\documentclass[preprint,12pt]{elsarticle}

\usepackage{epsfig}
\usepackage{amssymb}
\usepackage{amsmath}
\usepackage{hyperref}
\usepackage{float}
\usepackage{gensymb}
\usepackage{subcaption}
\usepackage{textcomp}  
\usepackage[utf8]{inputenc}
\DeclareUnicodeCharacter{2500}{-}
\usepackage{lineno}

\journal{.}

\begin{document}

\begin{frontmatter}

\title{Nanoscale Analysis of Surface Modifications on Silanized Glass: Wettability Alteration and Long\textendash Term Stability}

\author[1]{Mohammad Hossein Khoeini \corref{cor1}}
\author[1]{Gijs Wensink}
\author[2,3]{Tomislav Vukovic}
\author[1]{Ilja Krafft} 
\author[2,3]{Antje van der Net} 
\author[1,4,5]{Maja R\"ucker} 
\author[1,4]{Azahara Luna\textendash Triguero \corref{cor2}}

\affiliation[1]{organization={Eindhoven University of Technology, Department of Mechanical Engineering},country={the Netherlands}}
\affiliation[2]{organization={Norwegian University of Science and Technology},
country={Norway}}
\affiliation[3]{organization={Porelab Institute}, country={Norway}}
\affiliation[4]{organization={Eindhoven Institute for Renewable Energy Systems}, country={the Netherlands}}
\affiliation[5]{organization={Max Planck Institute for Polymer Research}, country={Germany}}

\cortext[cor1]{m.h.khoeini@tue.nl}
\cortext[cor2]{a.luna.triguero@tue.nl}

\begin{abstract} 
To investigate the effect of wettability on multiphase flow in porous media, hydrophilic glass surfaces are typically modified through a silanization process. This study examines the nanoscale chemical and structural modifications of glass bead surfaces treated with Surfasil, using inverse gas chromatography and atomic force microscopy. The results show that silanization reduces both specific and dispersive components of surface energy, indicating fewer polar groups and lower total energy, leading to decreased hydrophilicity compared to untreated glass beads. BET surface area measurements and AFM images reveal that the surface becomes progressively smoother with increased silanization.

Subsequently, this study assessed the stability and extent of surface modifications in silanized samples caused by adsorbed water during storage, using untreated glass beads as a reference. Untreated samples exhibit increases in surface roughness and polar groups, leading to marginal increase in surface energy and hydrophilicity. In contrast, the silanized samples show resistance to water adsorption, with only minor alterations in surface energy and structure, likely occurring in areas where the silanization coating was incomplete. The results suggest that humidity control is crucial during extended storage, as prolonged moisture exposure could still lead to surface modifications, even in silanized samples, potentially affecting wettability consistency in repeated experiments.
\end{abstract}

\begin{graphicalabstract}
\includegraphics[width=1\linewidth]{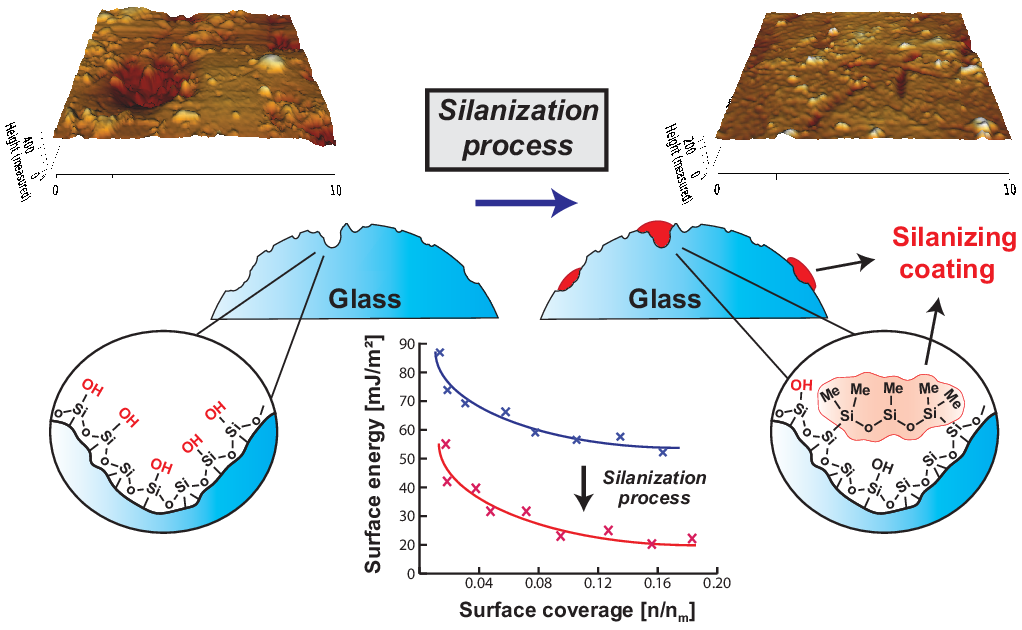}
\end{graphicalabstract}

\begin{highlights}
\item Silanization reduces polar groups and dispersive energy, increasing glass hydrophobicity.
\item AFM \& BET show smoother glass surfaces post-silanization, coating nanoscale roughness
\item Humidity induces nanoscale changes during storage of untreated and silanized samples
\item Model developed showing surface changes in glass during long-term storage conditions
\item Strict storage humidity control is vital for data reproducibility in future experiments
\end{highlights}


\end{frontmatter}

\section{Introduction}

Multiphase flow in porous media is critically important for various industrial and environmental applications, including underground hydrogen storage, oil recovery, groundwater remediation, ink spreading on paper, and carbon dioxide sequestration \cite{Heinemann2021EnablingChallenges,Morrow1986EffectRecovery,Mercer1990ARemediation,Boot-Handford2014CarbonUpdate,Wijshoff2018DropProcess}. Understanding and optimizing multiphase flow behavior are essential for enhancing the efficiency of these processes. Fluid displacement patterns in porous media are influenced by several factors, including flow rates, fluid properties (such as viscosity, density, and surface tension), and the characteristics of the porous medium (such as pore size, shape, and connectivity) \cite{Singh2018Capillary-DominatedMedia,Pavuluri2023InterplayMedia}. Among these factors, wettability of porous media plays a crucial role by determining relative permeability \cite{Armstrong2021MultiscaleMedia,Mascini2020Event-basedTomography}. The wettability of the porous medium, defined as the tendency of a solid to be in contact with one fluid rather than others, is typically expressed as the contact angle measured at the solid-fluid-fluid contact line through the denser phase.

Recent advancements in micro X-ray computed tomography (micro-CT) imaging have enabled visualization of multiphase flow behavior at the pore scale, providing a powerful tool to experimentally study the effects of wettability on fluid displacement patterns \cite{Berg2013Real-timeFlow,Shui2007MultiphaseInterfaces}. Glass is frequently used for fabricating porous bead packs and micro-models due to its high optical transparency, chemical inertness, and ease of molding into desired shapes \cite{Gaillard2018IsothermGlass,Bauer2019FunctionalizationChromatography}. The wettability of glass, which is inherently water-wet due to the presence of high hydroxyl group content on its surface \cite{Iglauer2014ContaminationMeasurements},  can be altered through a surface treatment method known as silanization \cite{Vukovic2023SystematicModification,Rueckriem2012InverseGlass.,Dey2016CleaningPerspective}. In this process, the hydrophilic groups on the surface are replaced by hydrophobic silanes and/or siloxanes groups \cite{Maharanwar2017AnalysisDepositions,Hoffmann2019Siloxane-functionalisedObjects}. By adjusting the treatment parameters of the silanization process, such as treatment time, type of silanizing agent, cleaning procedure before silanization, and solvent type, a wide range of contact angles on glass can be achieved \cite{Maharanwar2017AnalysisDepositions,McGovern1994RoleOctadecyltrichlorosilane,Borges-Munoz2019SilanizationP-aminophenyltrimethoxysilane,Cras1999ComparisonSilanization}. 

Although the influence of these reaction parameters is generally recognized, a standardized method for reliably achieving specific contact angles across various glass geometries, such as plates, microchips, and beads, was yet to be established \cite{Borges-Munoz2019SilanizationP-aminophenyltrimethoxysilane,Vukovic2023SystematicModification}. To address this gap, Vukovic et al. \cite{Vukovic2023SystematicModification} systematically explored the existing silanization treatment process using dichlorooctamethyltetrasiloxane (Surfasil) and proposed a refined procedure that achieves stable and reproducible contact angles across a broad range, which is applicable to various glass shapes. In their work, they initially investigated the influence of Surfasil solvent, treatment time, Surfasil-to-solvent ratio on the glass plates using sessile droplet method. By establishing the relationship between these parameters and optimizing them, while keeping temperature and cleaning procedure constant, they achieved stable and repeatable contact angles ranging from 20\degree (untreated glass) to 95\degree (Surfasil-coated glass) for the air-water-glass plate system. Subsequently, to evaluate the effectiveness and adaptability of the developed procedure to different glass geometries, they applied the treatment method to single glass bead, microchip, and bead packs. The wettability alteration on these geometries were then quantified using optical image analysis for microchips and single bead, and micro-CT image analysis for the bead packs. The results were consistent with those obtained for glass plates, confirming the effectiveness and transferability of the treatment method across different geometries.

Although the proposed treatment method appeared successful, their study primarily relied on contact angle measurements to evaluate changes in wettability. While this technique provides valuable macroscopic insights, it fails to reveal the underlying structural and chemical modifications at the nanoscale that drive wettability changes \cite{Bendada2016SurfaceSurfactant}. Specifically, the extent of chemical modification is directly tied to the degree of surface coverage by the silanizing agent, while the structural modification depends on how the silanized groups are arranged on the surface \cite{Bauer2019FunctionalizationChromatography}, which contact angle measurements fail to identify independently. To address this limitation, our study aims to investigate these surface alterations using inverse gas chromatography (IGC) and atomic force microscopy (AFM). AFM is a molecular-scale imaging technique that utilizes a cantilever with a sharp tip to mechanically scan specific regions of the sample surface, generating high-resolution topographical maps by measuring forces between the tip and the surface. This allows us to directly observe nanoscale structural modifications. In contrast, IGC employs gas molecules as probes to characterize the physicochemical properties of the surface, such as surface area and energy distribution. The advantage of IGC over AFM lies in its ability to provide an average measurement of the entire surface, due to the gaseous nature of the probe, rather than being confined to the specific scanned areas \cite{Khoeini2025AMedia}. Together, these advanced analytical techniques offer a comprehensive characterization of the nanoscale structural and chemical changes induced by the Surfasil silanization treatment, providing a deeper understanding of the mechanisms underlying wettability modifications.

\begin{figure}[!ht]
    \centering
    \includegraphics[width=1\linewidth]{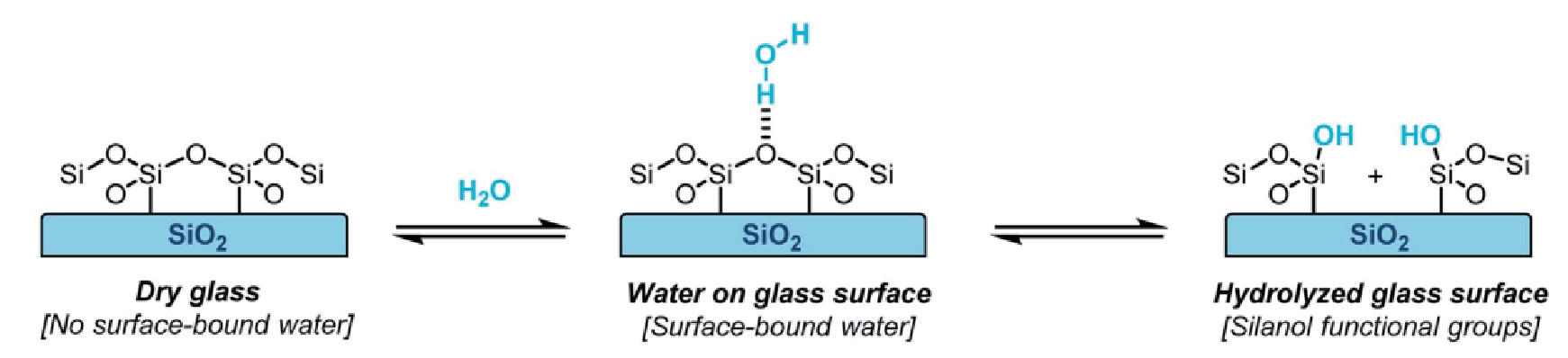}
    \caption{Simplified illustration of water adhesion on a pure silica glass surface and hydroxylation of siloxane groups. Reproduced from \cite{Nielsen2022VesselPhenomenon} under the Creative Commons Attribution 3.0 License (CC BY 3.0).}
    \label{fig:Figure_1}
\end{figure}

Furthermore, the stability of the silanized coating on modified glass is crucial for ensuring consistent wettability during multiphase flow experiments and, more importantly, for the reproducibility of results when samples are stored. To examine the coating's stability, Vukovic et al. \cite{Vukovic2023SystematicModification} stored the treated glass plate samples, which initially had a contact angle approximately 95\degree for water-air-glass, in both air and distilled water under atmospheric conditions. While the contact angle measurements for the samples stored in distilled water showed a decrease of 2\degree over 4 days, the samples stored in air remained stable within the error margin over 143 days, leading to the conclusion that the coating is stable when stored in air, consistent with results reported in other studies \cite{Gaillard2018IsothermGlass,Wei1993WettingOil,Menawat1984ControlStability}. Despite this conclusion, from a chemical standpoint, the adhesion of water to the untreated glass surface, even in the form of adsorbed water from the air, can initiate dissolution, ion-exchange, and hydroxylation reactions (as shown in \autoref{fig:Figure_1}). These reactions result in chemical alterations, such as an increase in hydrophilic sites like hydroxyl groups, and structural changes, such as increased surface roughness \cite{Yeon2023HydroxylationSimulations,Nielsen2022VesselPhenomenon}. The extent and rate of these reactions strongly depend on the relative humidity of the air, temperature, glass composition, and duration of storage \cite{Gin2021AqueousPerspectives}. 

The characterization method used in the Vukovic et al. study \cite{Vukovic2023SystematicModification}, specifically contact angle measurement, was unable to determine whether chemical and structural modifications occurred at the nanoscale, as phenomena at this scale surpass the resolution capability of the employed method. These modifications, even at a nanometer scale, can impact the distribution of fluids' thin films on the surface. Such thin films are crucial in governing multiphase flow behavior, particularly near the saturation endpoint of the wetting fluid \cite{Armstrong2021MultiscaleMedia,Wensink2024In-situRocks}. As a result, the chemical and structural modifications caused by the reaction of adsorbed water from air with stored samples can influence multiphase flow patterns, resulting in inconsistencies in repeated experiments. To address this gap, our study employs IGC and AFM to investigate these nanoscale modifications, thereby enhancing the reliability of wettability consistency in repeated experiments.

\section{Methods and Theories}

\subsection{Glass beads silanization procedure}
\label{sub_sec:GB_Sila_Procedure}

To investigate the structural and chemical modifications of glass through the silanization procedure proposed by Vukovic et al. \cite{Vukovic2023SystematicModification}, and to assess the degree of these modifications when the treated samples are stored in air, soda lime glass beads of 1 mm and 2 mm in size were purchased from Karl Hecht Assistent. Dichlorooctamethyltetrasiloxane (Surfasil) for the silanization process was obtained from Thermo Scientific. To prepare the desired concentration of the Surfasil agent, n-heptane ($>$99\%), as recommended in the study by Vukovic et al., was purchased from VWR Chemicals. The wettability alterations on the glass bead samples through silanization were performed following the steps outlined in Vukovic et al..

\textbf{Cleaning}: In the first step, the glass beads were cleaned using a miscible rinsing sequence of toluene, methanol, and acetone, with each rinsing sequence lasting at least 30 seconds. The selected miscible solvents and their sequence cover a wide range of polarities, enabling the dissolution of surface contaminants and the removal of preceding solvents. Following the rinsing process, the samples were dried using a nitrogen gun and then placed in an oven at 80 \degree C for two hours to ensure complete removal of any residual fluids. To prevent re-contamination, all samples were handled exclusively with tweezers. 

\textbf{Surfasil Coating}: The coating was applied by submerging the cleaned glass beads into a beaker containing diluted Surfasil in n-heptane at room temperature (23 \degree C). After 180 seconds, the glass beads were removed from beaker using a sieve and immediately rinsed with pure n-heptane followed by methanol, for approximately 30 seconds each, to stop the silanization reaction and prevent the reaction of Surfasil with water. Subsequently, the samples were placed in an oven at 80 \degree C for 2 hours to evaporate the solvents and finalize the cross-linking of the coating. 

\begin{figure}[!ht]
    \centering
    \includegraphics[width=0.65\linewidth]{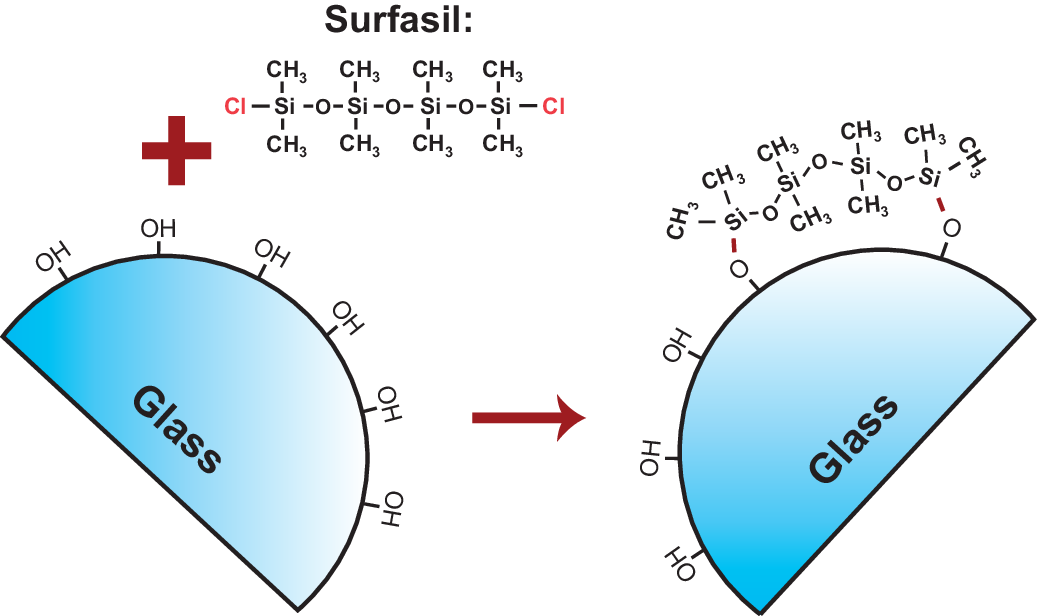}
    \caption{Mechanism of the silanization reaction on the glass surface.}
    \label{fig:Figure_2}
\end{figure}

The hydrophilicity of a glass surface is primarily attributed to the presence of polar groups, such as silanol groups \cite{Dey2016CleaningPerspective}. During the silanization process, the chloride groups of Surfasil react with the silanol groups on the surface, producing hydrogen chloride and siloxane with methyl groups (as shown in \autoref{fig:Figure_2}). The packing density of the siloxane coating directly controls the wettability alteration \cite{McGovern1994RoleOctadecyltrichlorosilane}. To achieve a wide range of wettability alterations, the glass bead samples of both sizes were treated with different Surfasil-to-solvent ratios (VR): 0.0002, 0.001, and 0.1. For comparison, all measurements of the following sections were also performed on untreated glass beads to serve as a reference.

\subsection{Wettability analysis by contact angle measurements}
\label{sub_sec:CA_Wet_analysis}

The contact angle measurements for the single bead of different samples (described in \autoref{sub_sec:GB_Sila_Procedure}) in distilled water-air system were performed by the Sessile droplet method performed with a Kruss DSA100S drop shape analyzer at room temperature (23 \degree C). The Kruss DSA100S device consists of a high-power monochromatic LED illumination source, a camera system with 20 $\mu$m optical resolution, and a software-controlled syringe system with a resolution of 0.1 $\mu$L for dosing liquid. The procedure for these measurements has been comprehensively detailed in a prior publication \cite{Vukovic2023SystematicModification}, and a brief summary of the steps is provided below.

To determine the contact angle, the following steps were taken: first, a droplet of distilled water was placed on a glass plate, and then a glass bead (2mm) was carefully placed in the middle of the droplet using tweezers. The device's camera system was then used to obtain high-quality images of the single bead-water system. Imaging continued until all of the water evaporated, leaving only the single bead on the glass plate.

Subsequently, image processing was conducted using ImageJ software. The image of the single bead on the plate surface, after all the water had evaporated (dry image), was overlapped with the initial image taken when the single bead was placed on the water droplet (wet image). By overlapping the wet and dry images, the three-phase contact point and contact angle were determined. To ensure accuracy, the experiments were repeated using three different single beads for each sample type, and contact angle measurements were taken at three different locations on the glass plate for each bead.

\subsection{Inverse Gas Chromatography (IGC)}

\begin{figure}[!ht]
    \centering
    \includegraphics[width=1\linewidth]{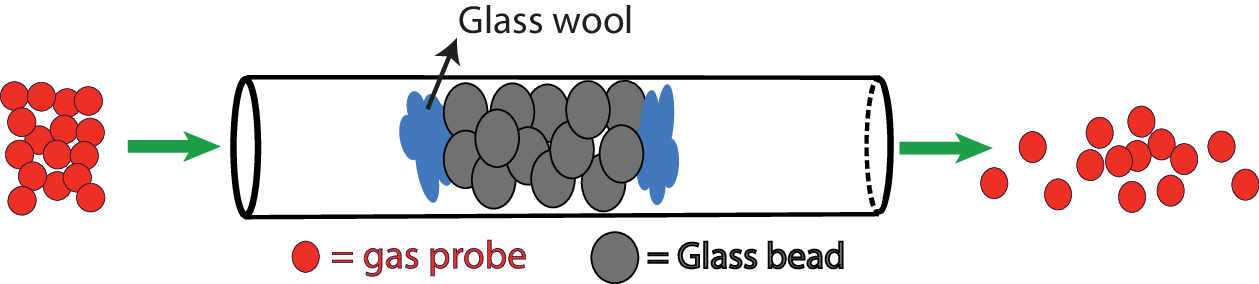}
    \caption{Schematic of an Inverse Gas Chromatography column packed with glass beads.}
    \label{fig:Figure_3}
\end{figure}

IGC is an accurate, sensitive, and relatively fast method for exploring the physiochemical properties of the surfaces of non-volatile materials at the molecular scale \cite{Ho2013ASolids,Voelkel2004InverseSurface,Khoeini2025AMedia}. In this method, a pulse of a single known gas, referred to as a probe, is injected into a column packed with the sample of interest (as showin in \autoref{fig:Figure_3}). The injected molecules into the column, depending on their chemical nature, interact with the surface through dispersive, known as Lifshitz-van der Waals interactions, and/or specific forces, which encompass all other types of interactions. The time it takes for different probes to travel through the column, known as retention time, is used to determine the thermodynamic properties of the sample surface, such as surface area and surface energy distribution. The changes in surface area measured via IGC, together with AFM images, enable us to study the structural changes resulting from the silanization process or reactions with adsorbed water. Additionally, variations in surface energy distribution offer insights into the chemical alterations of the surface. By combining these IGC measurements with AFM data, a comprehensive understanding of both the structural and chemical modifications in the glass beads can be achieved. The following subsections will first explore the theoretical framework of IGC and then describe the experimental procedures utilized in this study.

\subsubsection{Theoretical Aspects of Inverse Gas Chromatography}

The retention time of the probe ($t_{R}$), which reflects the strength of interactions between the probe and the surface, is the primary result of an IGC measurement. The net retention volume, $V_{n}$, defined as the volume of carrier gas required to elute the injected probe molecules, can be calculated from the retention time using Eq.\ref{eq:EQ_1}: 

\begin{equation}
\label{eq:EQ_1}
V_{n} = \frac{j}{m}.F.(t_{R}-t_{0}).\frac{T}{273.15}
\end{equation}

Where $T$ is the column temperature in Kelvin (K), F is carrier gas exit flow rate at 1 atm and 273.15 K, $t_{R}$ is retention time of injected probe, $t_{0}$ (dead time) is the time required for the probe molecule to travel through the column without any interaction (usually methane), m is the sample mass in the column, and j is the James–Martin correction for taking into account the compressibility of the carrier gas under the effect of pressure drop ($\Delta P$) along the column and also variation in packing density of the samples within the column bed \cite{Ho2013ASolids}. The James–Martin correction factor is defined as:

\begin{equation}
\label{eq:EQ_2}
j = \frac{3}{2}.[\frac{(1+\frac{\Delta P}{P_{atm}})^2-1)}{(1+\frac{\Delta P}{P_{atm}})^3-1)}]
\end{equation}

The net retention volume can be correlated to the first derivative of the isotherm derived from the mass balance equation of the IGC column, which is valid under the following conditions \cite{Seidel-Morgenstern2020ModelingProcesses,Dorris1980AdsorptionFibers,Conder1979PhysicochemicalChromatography}: (1) the probe is retained primarily through surface adsorption on the solid surface, with negligible adsorption on the column wall; (2) the system is isothermal; (3) the influence of axial dispersion is ignored; (4) external and internal mass transfer coefficients are very high; and (5) local equilibrium is maintained, meaning the rates of adsorption and desorption are very fast. The equation is given by: 

\begin{equation}
\label{eq:EQ_3}
(\frac{\partial N}{\partial P})_{L,t_{r}} = \frac{V_{n}}{R.T}
\end{equation}

where N is the number of adsorbed probe molecules, P is the pressure of the probe at the column outlet, which is directly related to the detector signal, R is the ideal gas constant, and T is the temperature of the column. 

Depending on the concentration of the probe molecules injected into the column, IGC measurements can be conducted in two distinct regimes: infinite dilution and finite concentration. In the infinite dilution regime (IGC-ID), a very low concentration of the probe (typically less than $3\%$ of its partial pressure) is injected, resulting in minimal surface coverage, known as the Henry's region. In this regime, the adsorption of probe molecules is linearly dependent on their concentration, and interactions between probe molecules are negligible. Consequently, the retention time primarily reflects the interactions between the probe and the solid surface. Beyond the Henry's region, the system enters the finite concentration regime (IGC-FC), where higher concentrations of the probe are used, leading to the formation of a probe monolayer on the solid surface. In the following sections, the theoretical frameworks for the determination of specific surface area and surface energy distribution are presented \cite{Ho2013ASolids,Thielmann2007DeterminationLactose,Mohammadi-Jam2014InverseReview}.

\paragraph{\textbf{Surface Energy Concepts in IGC}}

At infinite dilution conditions, interactions between probe molecules are negligible, and the interactions between the probe and the solid surface are dominant. In this region, by incorporating the linearity of the isotherm (Eq.\ref{eq:EQ_3}) into the Gibbs adsorption isotherm, the standard free energy of adsorption ($\Delta G_{ad}^{\,0}$), or desorption ($\Delta G_{de}^{\,0}$), per mole of molecules can be related to the retention volume via \cite{Dorris1980AdsorptionFibers,Schultz1987TheComposites}:

\begin{equation}
\label{eq:EQ_4}
\Delta G_{de}^{\,0} = -\Delta G_{ad}^{\,0} = RT\: ln(\frac{V_{n}.\,P_{0}}{S.\,\pi_{0}})=RT\:ln(V_{n})+C
\end{equation}

Where S is specific surface area of the probe, $\pi_{0}$ is the spreading pressure of the chosen surface standard state, and $P_{0}$ is the corresponding partial pressure of the probe at the reference state. The choice of reference state is somewhat arbitrary; however, two reference states proposed by Kemball and Rideal ($P_{0}=1.013\,\times 10^{5}\:Pa$ and $\pi_{0}=6.08\,\times 10^{-5}\:Nm^{-1}$) or De Boer ($P_{0}=1.013\,\times 10^{5}\:Pa$ and $\pi_{0}=3.38\,\times 10^{-4}\:Nm^{-1}$) are widely used \cite{Dorris1980AdsorptionFibers,Schultz1987TheComposites,Gholami2020SurfaceReview}. The standard Gibbs energy of sorption can then be related to the energy of adhesion between the probe and the solid, per unit surface area, by:

\begin{equation}
\label{eq:EQ_5}
\Delta G_{ad}^{\,0} = N_A.\,a_m.\,W_A
\end{equation}

Where $N_A$ is Avagardo's number, and $a_m$ is the cross-sectional area of a single adsorbed molecule. 
The relation between retention volume and the work of adhesion can be established using Eq.\ref{eq:EQ_4} into Eq.\ref{eq:EQ_5}:

\begin{equation}
\label{eq:EQ_6}
RT\:ln(V_{n})= N_A.\,a_m.\,W_A\,+C
\end{equation}

Surface energy quantifies the attractive intermolecular forces between the surface and the probe. As previously mentioned, probe molecules interact with the solid surface through various intermolecular forces, depending on their chemical nature. Work by Fowkes on surface adsorption \cite{Fowkes1981Acid-BaseAdhesion,Fowkes1981Acid-BaseAdhesion,Fowkes1978Acid-BaseAdsorption,Fowkes1987RoleAdhesion}, later referred to as the surface tension component theory by some researchers \cite{Lee1993RolesWettability}, concluded that only similar types of forces interact across an interface. Consequently, the total surface energy, the total work of adhesion, and the total Gibbs free energy of sorption can be divided into several independent, or partially independent components, each representing a different type of intermolecular interaction. These partly individual components are later grouped into two main parts: dispersive and specific (as shown in Eq.\ref{eq:EQ_7}) \cite{VanOss1988AdditiveAngles,vanOss1987MonopolarSurfaces}. The dispersive component includes Lifshitz-van der Waals interactions, while the specific component encompasses all remaining types of interactions, such as hydrogen-bonding, acid-base, metallic, and magnetic interactions \cite{Goss1997ConsiderationsCoefficients}. In the absence of electrostatic, magnetic, or metallic interactions, acid-base interactions become the dominant factor \cite{Voelkel2009InverseData,Balard2000DeterminationLimitations}. Consequently, the specific component is often referred to as acid-base interaction.

\begin{equation}
\label{eq:EQ_7}
\begin{split}
W_{A}^{t} & =W_{A}^{D}+W_{A}^{SP}\\ 
\Delta G_{ad}^{\,t} & =\Delta G_{ad}^{\,D}+\Delta G_{ad}^{\,SP}\\ 
\gamma_{ad}^{\,t} & =\gamma_{ad}^{\,D}+\gamma_{ad}^{\,SP}
\end{split}
\end{equation}

Where $W_A$, and $\gamma_{ad}$ represent the work of adhesion, and surface energy, respectively. The superscripts $t$, $D$, and $SP$ correspond to the total, dispersive, and specific components, respectively.

\subparagraph{Dispersive component of surface energy}
\label{Par:Dispersive_Component}
When using non-polar probes that interact exclusively through dispersive forces with the surface, the contribution of the specific component to the total work of adhesion is negligible ($W_{A}^{SP}=0$). Fowkes demonstrated that the dispersive component of the work of adhesion can be calculated using the geometric mean method \cite{Fowkes1964ATTRACTIVEINTERFACES}:

\begin{equation}
\label{eq:EQ_8}
W_{ad}^{D}=2\sqrt{\gamma_{P}^{D}.\,\gamma_{S}^{D}}
\end{equation}

Where the subscripts $P$ and $S$ refer to the probe and solid, respectively. By incorporating Eq.\ref{eq:EQ_8} into Eq.\ref{eq:EQ_6}, the relationship between the retention volume and the dispersive component of surface energy for the non-polar probes can be derived as follows:

\begin{equation}
\label{eq:EQ_9}
RT\:ln(V_{n})= N_A.\,a_m.\,2.\,\sqrt{\gamma_{P}^{D}.\,\gamma_{S}^{D}}\,+C
\end{equation}

\begin{figure}[!ht]
    \centering
    \includegraphics[width=0.63\linewidth]{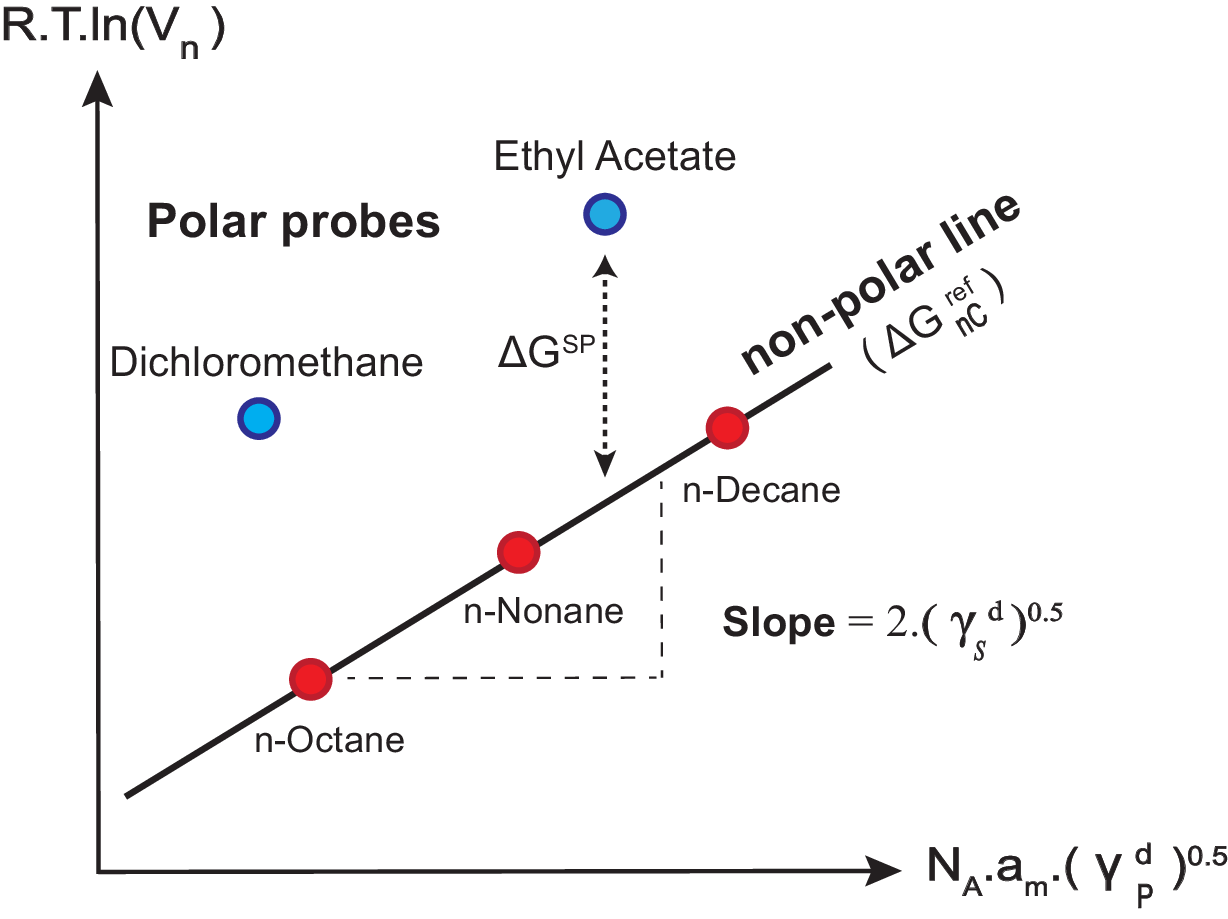}
    \caption{Schematic illustrating the determination of dispersive component of surface free energy (slope of non-polar line) and Gibbs energy of sorption ($\Delta G^{SP}$) from the Schultz method.}
    \label{fig:Figure_4}
\end{figure}

Schultz et al. \cite{Schultz1987TheComposites} reported that the dispersive surface energy of a solid can be obtained using series of n-alkanes by plotting $RT\:ln(V_{n})$ as function of $N_A.a_m.(\gamma_{P}^{D})^{0.5}$. The dispersive component of surface energy ($\gamma_{S}^{D}$) can then be calculated from the slop of the linear regression (as shown in \autoref{fig:Figure_4}).

\subparagraph{Specific component of surface energy}
\label{Par:Specific_Component}
With the use of polar probes, the total Gibbs free energy of sorption and total work of adhesion are now composed of both dispersive and specific components. Since retention time of polar probe, which reflects the combined effect of both interactions, is the only measured parameter, it is necessary to separate and evaluate the contribution of each component. Typically, the specific component of the Gibbs energy of sorption ($\Delta G^{SP}$) is determined as difference between the total Gibbs energy of sorption of polar probe, and Gibbs energy of hypothetical n-alkane ($\Delta G_{nC}^{\,ref}$), which has the same properties as polar probe for a given $N_A.a_{m}.\,\sqrt{\gamma_{P}^{D}}$ in Schultz method (shown in \autoref{fig:Figure_4}). Van Oss et al. \cite{VanOss1988AdditiveAngles}, based on the concept that the specific component depends on the availability of sites capable of undergoing specific types of interactions, proposed sub-dividing this component into acid (or electron-acceptor $\gamma_{S}^{+}$) and basic (or electron-donor, $\gamma_{S}^{-}$) parameters as follow:

\begin{equation}
\label{eq:EQ_10}
\begin{split}
\Delta G_{ad}^{\,SP} = 2.\,&(\sqrt(\gamma_{S}^{+}\,\gamma_{P}^{-})+\sqrt(\gamma_{S}^{-}\,\gamma_{P}^{+}))\\ 
\gamma_{S}^{\,SP} &= 2.\,\sqrt(\gamma_{S}^{-}.\gamma_{S}^{+})
\end{split}
\end{equation}

To determine the acidic and basic parameters of the specific component, two monopolar probes, which interact solely through either acidic or basic interactions with the surface, can be employed. Finally, according to Owens and Wendt \cite{Owens1969EstimationPolymers}, the work of adhesion between a solid and a polar probe can be determined from the summation of the geometric mean terms of dispersive and specific surface energy components, as shown by:
\begin{equation}
\label{eq:EQ_11}
W_{A}^{\,t} = 2.\sqrt(\gamma_{S}^{D}\,\gamma_{P}^{D})\,+\,2.\sqrt(\gamma_{S}^{SP}\,\gamma_{P}^{SP})
\end{equation}

\subparagraph{Surface energy distribution}
\label{Par:SE_distribution}
Injected probe molecules typically first interact with the most energetic sites on the surface. Once these high-energy sites are occupied, subsequent probe molecules interact with progressively lower energy sites until the interactions between probe and surface sites become comparable to those between probe molecules, entering the multilayer adsorption regime. In the infinite dilution mode of IGC, the use of small quantities of probe molecules preferentially samples these high-energy sites, leading to a biased representation that does not accurately reflect the entire surface's energy distribution \cite{vanAsten2000SurfaceChromatography,Buckton2007TheChromatography}. To account for all high and low energy sites, measurements in both IGC-FC and IGC-ID modes should be used. 

Thielmann et al. \cite{Thielmann2007DeterminationLactose,Ho2013ASolids} developed a methodology that utilizes both modes to measure surface energy heterogeneity. This approach involves multiple injections, with gradually increasing concentrations, of a series of n-alkanes and polar probes to determine the dispersive and specific surface energy distributions. The first step in this methodology is to determine the adsorption isotherms for all probes from their injections. Each injection of a probe must be correlated with its corresponding surface coverage ($N/n_m$) to obtain the distribution. By assuming that the isotherms of n-alkanes follow Type II or V, according to IUPAC classification \cite{Thommes2015PhysisorptionReport}, the BET model can then be applied to one of n-alkane isotherm to determine the specific surface area ($S_{BET}$). This specific surface area can also be determined separately using an alternative technique, such as nitrogen adsorption at 77K. The next step is to determine the monolayer capacity ($n_m$) for each probe from the BET surface area as follows:

\begin{equation}
\label{eq:EQ_12}
S_{BET} = N_A.\, a_m.\,n_m
\end{equation}

Once the monolayer capacity for each probe is known, the retention volumes of probes at each injection can be expressed as a function of their corresponding surface coverage. The subsequent step involves calculating the dispersive component from n-alkanes injections using the Schultz approach (as detailed in the \autoref{Par:Dispersive_Component}), and the specific component from polar probe injections (as explained in the \autoref{Par:Specific_Component}), for each surface coverage. By calculating $\gamma_S^D$ and $\gamma_S^{SP}$ across varying surface coverages, a distribution of dispersive and polar surface energies is obtained.

From this approach, the dispersive and specific surface energy components are obtained as a function of surface coverage, providing valuable insight into the surface's energy heterogeneity. At low surface coverage, where the infinite dilution (IGC-ID) mode is dominant, the measurements reflect the energy of high-energy surface sites, such as cracks and narrow pores. In contrast, at high surface coverage, the reported surface energy represents an average of most surface sites, including both high- and low-energy regions \cite{Ho2013ASolids,Thielmann2007DeterminationLactose,Yla-Maihaniemi2008InverseParticulates,Khoeini2025AMedia}. By comparing the surface energies at low coverage across different samples, one can assess changes in the high-energy sites, such as the extent to which cracks or pores have been chemically modified or shielded. Similarly, comparisons at high surface coverage provide an indication of how the overall surface properties, including the majority of the surface area, have changed as a result of treatment or modification.

\paragraph{\textbf{Specific surface area}}
The surface area of a solid is most commonly determined by applying the Brunauer–Emmett–Teller (BET) theory to the adsorption isotherms of probes \cite{Thommes2015PhysisorptionReport}. For this theory to be applicable, the adsorption isotherm of the probe must correspond to Type II or Type V classifications according to IUPAC, where the binding affinity between the probe and the solid is greater than that between probe molecules, thereby resulting in monolayer/multilayer formation mechanism. In the context of IGC, n-alkanes are typically used as probes for surface area determination due to their exclusive interaction through dispersive forces. As previously mentioned, the initial step in determining the surface area is to obtain the adsorption isotherm. This is achieved by integrating over Eq.\ref{eq:EQ_2} as follow:

\begin{equation}
\label{eq:EQ_13}
N = \frac{1}{R.T}\int_{0}^{P}\,V_n.\,dp
\end{equation}

To obtain the adsorbed amount, a relationship between the retention time and the pressure of the probe must be established. This can be achieved by injecting a series of pulses with different concentrations and correlating the retention time of each pulse with its corresponding pressure at the end of the column. Once the adsorption isotherm is determined, the BET equation \cite{Brunauer1938AdsorptionLayers} can be applied:

\begin{equation}
\label{eq:EQ_14}
\frac{P}{N.(P_0-P)}=\frac{1}{n_m.C}+\frac{C-1}{n_m.\,C}.\,\frac{P}{P_0}
\end{equation}

Where C is constant. Once the monolayer capacity, $n_m$, is determined, the specific surface area can be calculated from Eq.\ref{eq:EQ_12}.

\subsubsection{Experimental Procedure Using IGC}
\label{Sub_sub:IGC_exp_protocol}
In this study, inverse gas chromatography experiments were carried out using the IGC-SEA system from Surface Measurement Systems Ltd., London, UK. The glass bead samples, both untreated and treated via the procedure detailed in \autoref{sub_sec:GB_Sila_Procedure}) were packed into standard columns (30 $\times$ 0.3 cm ID) using tapping machine from the same supplier and secured in place with silanized glass wool. 

Approximately 1.5-2 grams of glass bead samples were packed into the columns, which were then conditioned in situ for 3 hours at 60 \degree C to remove any physisorbed impurities. All IGC measurements were performed at 30 \degree C, with flow rate of 10 ml/min, using nitrogen as the carrier gas, and methane as the non-retained probe. To measure the dispersive component of the surface energy of samples, n-Octane (\textgreater99.7\%), n-Nonane (\textgreater99\%), and n-Decane (\textgreater99\%) were used. For determination of specific component, Dichloromethane (\textgreater99.9\%), as acidic probe, and Ethyl acetate (\textgreater99.9\%), as basic probe, were used. All chemicals were purchased from Sigma-Aldrich. The BET surface area reported in this study was specifically calculated from the n-octane adsorption isotherm.

To investigate the chemical and structural modifications resulting from Surfasil silanization treatment, IGC measurements were performed on sample columns of untreated and treated glass beads (VR-0.0002, VR-0.001, VR-0.1). To verify the accuracy of BET surface area determination, two additional identical columns were prepared for the untreated glass beads. For the treated samples, due to the limited quantity available, BET values were determined from a single column. 

To study the surface modification caused by the reaction of adsorbed water from air during storage, two column samples were selected: one untreated (reference) and one VR-0.1 treated (most hydrophobic). When glass is in contact with humid air, the mechanisms of surface alteration differ from those in contact with liquid water. In humid air, the hydration process involves the redistribution of elements within the thin film of adsorbed water, rather than their release into the liquid, except for elements that can volatilize \cite{Majerus2020GlassManagement}. The number of adsorbed water molecule layers on the glass surface is primarily determined by temperature and relative humidity, with higher levels of both generally leading to an increased rate of hydration \cite{Asay2005EvolutionTemperature}. To simulate long-term storage conditions in an accelerated manner, the samples were exposed to 90\% relative humidity, with the remaining 10\% being nitrogen, for 20 days at 30\degree C. After exposure, the samples were dried at 60\degree C for 3 hours under a pure nitrogen stream to remove any remaining adsorbed water prior to IGC measurements. During this drying process, dissolved elements that were redistributed in the thin film but did not react elsewhere may precipitate at other locations on the surface. The entire process, including humidity exposure, drying, and measurements, was conducted within the IGC device.

\subsection{Atomic Force Microscopy (AFM)}

To investigate the topological alterations of the glass bead surfaces due to the Surfasil silanization process and subsequent water reactions, AFM imaging was conducted. AFM is an advanced imaging technique commonly used to obtain topographic images of surfaces at the nanometer scale. This technique operates by mechanically imaging a surface using a sharp tip mounted on the end of a microscopic cantilever, which performs raster-scanning over the surface. The tip interacts with the surface through various intermolecular forces, causing the cantilever to bend, which is then detected by a laser and a sensor \cite{Savulescu2021AtomicSurface}.

In this study, AFM imaging was performed on 2 mm diameter samples of untreated and silanized glass beads to investigate nanoscale surface modifications induced by the silanization process. To assess the modifications caused by adsorbed water during storage, AFM imaging was also conducted on 1 mm diameter samples of untreated and VR-0.1 treated beads, both before and after exposure to humidity. A JPK instrument bioAFM with a PPP-NCHAuD probe from NANOSENSORS was used for the imaging. Height images were obtained in Quantitative-imaging (QI) mode by measuring 128 $\times$ 128 force-distance curves over a 10 $\mu$ m x 10 $\mu$m area, with an approach height of 1 $\mu$m and a velocity of 100 $\mu$m/s. From these height images, histograms were generated to represent the distribution of surface heights, with frequencies plotted relative to the modal height, or the most frequently observed height.

\section{Results and Discussion}

This section is divided into two parts: the first part provides a detailed discussion of the findings on the structural and chemical modifications of glass bead surfaces resulting from the silanization treatment. The second part examines the structural and chemical modifications of untreated and treated glass beads caused by the reaction of adsorbed water from air during storage.

\subsection{Surface Modifications from Silanization Treatment} 

\begin{figure}[!ht]
        \centering
        \begin{subfigure}[t]{0.48\textwidth}
            \centering
            \includegraphics[width=1\linewidth]{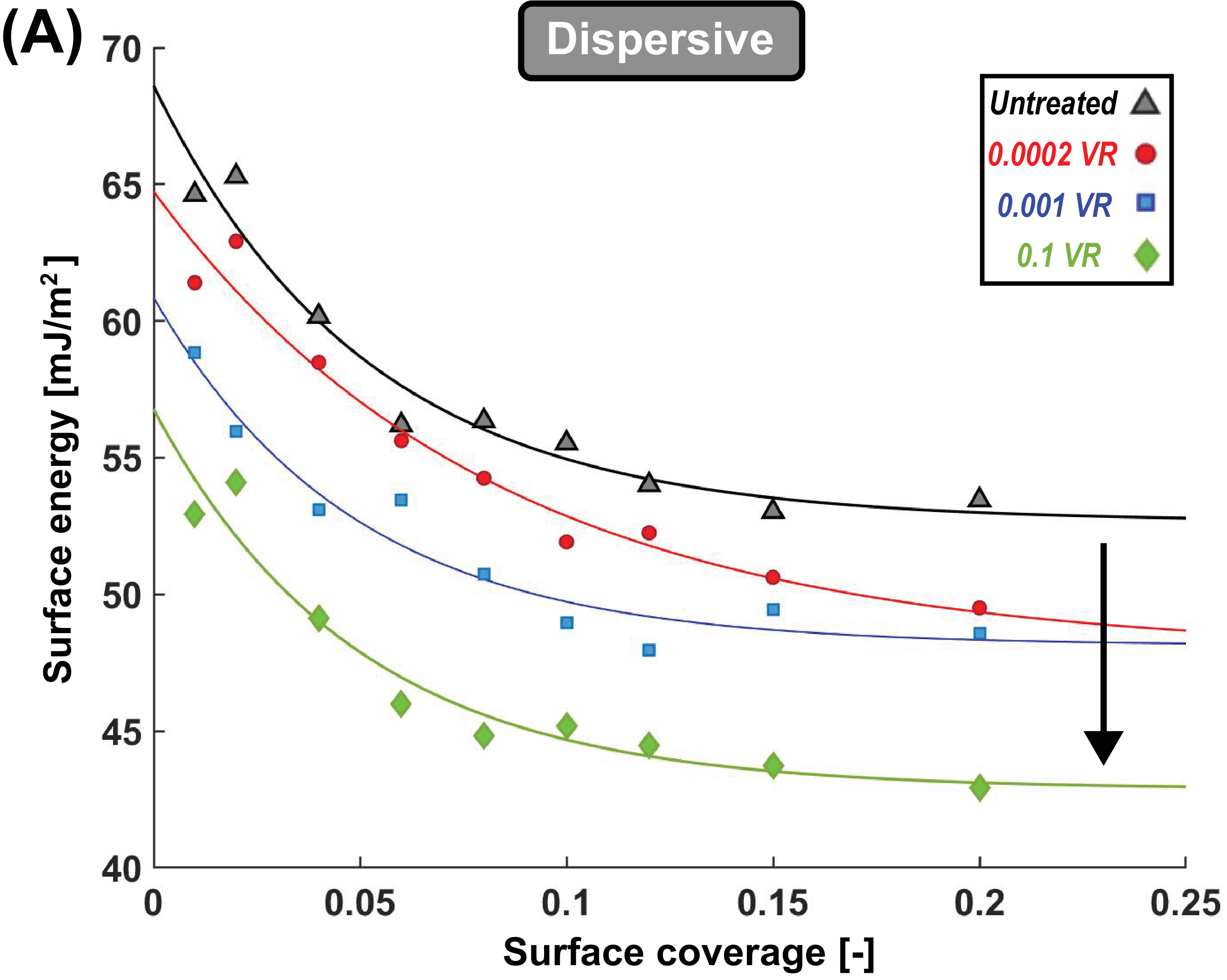}
        \end{subfigure}
        \hfill
        \begin{subfigure}[t]{0.48\textwidth}
            \centering
            \includegraphics[width=1\linewidth]{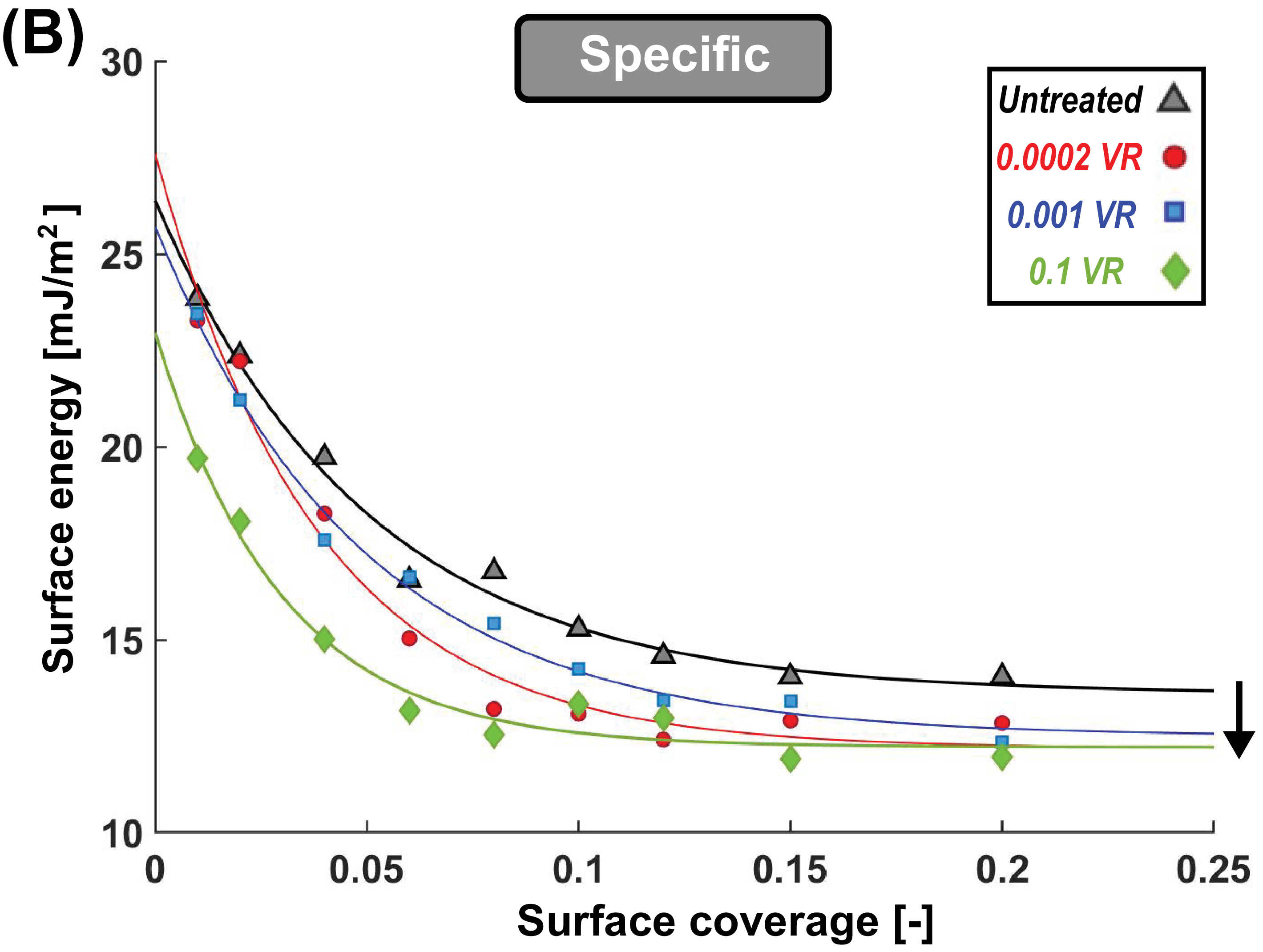}
        \end{subfigure}      
    \caption{Dispersive and specific components of surface energy for untreated and treated glass beads.}
    \label{fig:Figure_5}
\end{figure}

The wettability of material surfaces is typically quantified by measuring the contact angle. The contact angles, measured using the sessile droplet method for the distilled-air-sample system as detailed in \autoref{sub_sec:CA_Wet_analysis}, are $ 23.95 \pm 1.25\degree$ for untreated glass beads, $49.78 \pm 2.35\degree$ for VR-0.0002 treated sample, $64.2 \pm 2.04\degree$ for the VR-0.001 treated sample, and $102.51 \pm 1.71\degree$ for the VR-0.1 treated glass beads \cite{Vukovic2023SystematicModification}. These measurements indicate a trend towards a more hydrophobic state with increased treatment. While contact angle measurements provide valuable macroscopic data, they do not offer insights into nanoscale chemical and structural changes. To assess surface chemical modifications at nanoscale, surface energy of samples were obtained using IGC, as shown in \autoref{fig:Figure_5}. The results show that both the dispersive and specific components of surface energy decrease with increased silanization. According to Young's equation, a reduction in total surface energy, which is the sum of the dispersive and specific components, should lead to an increase in the contact angle. Therefore, the observed trends in surface energy components are consistent with the measured contact angle data.

As previously indicated in \autoref{sub_sec:GB_Sila_Procedure}, the hydrophilicity of a glass surface is primarily due to the abundant presence of polar groups, such as silanol groups. The primary objective of silanization process is to reduce this hydrophilicity by substituting these polar groups with a silanizing reagent (Surfasil, in this instance) \cite{Dey2016CleaningPerspective}, as depicted in \autoref{fig:Figure_2}. The reduction in the specific component of surface energy in the treated samples, as shown in Figure \ref{fig:Figure_5}.b, indicates that the silanization process effectively targeted the specific groups of surface, a trend similarly observed in previous studies \cite{Rueckriem2012InverseGlass.,Mader-Arndt2014SingleParticles,Bauer2019FunctionalizationChromatography}.

\begin{figure}[!ht]
        \centering
        \begin{subfigure}[t]{0.48\textwidth}
            \centering
            \includegraphics[width=1\linewidth]{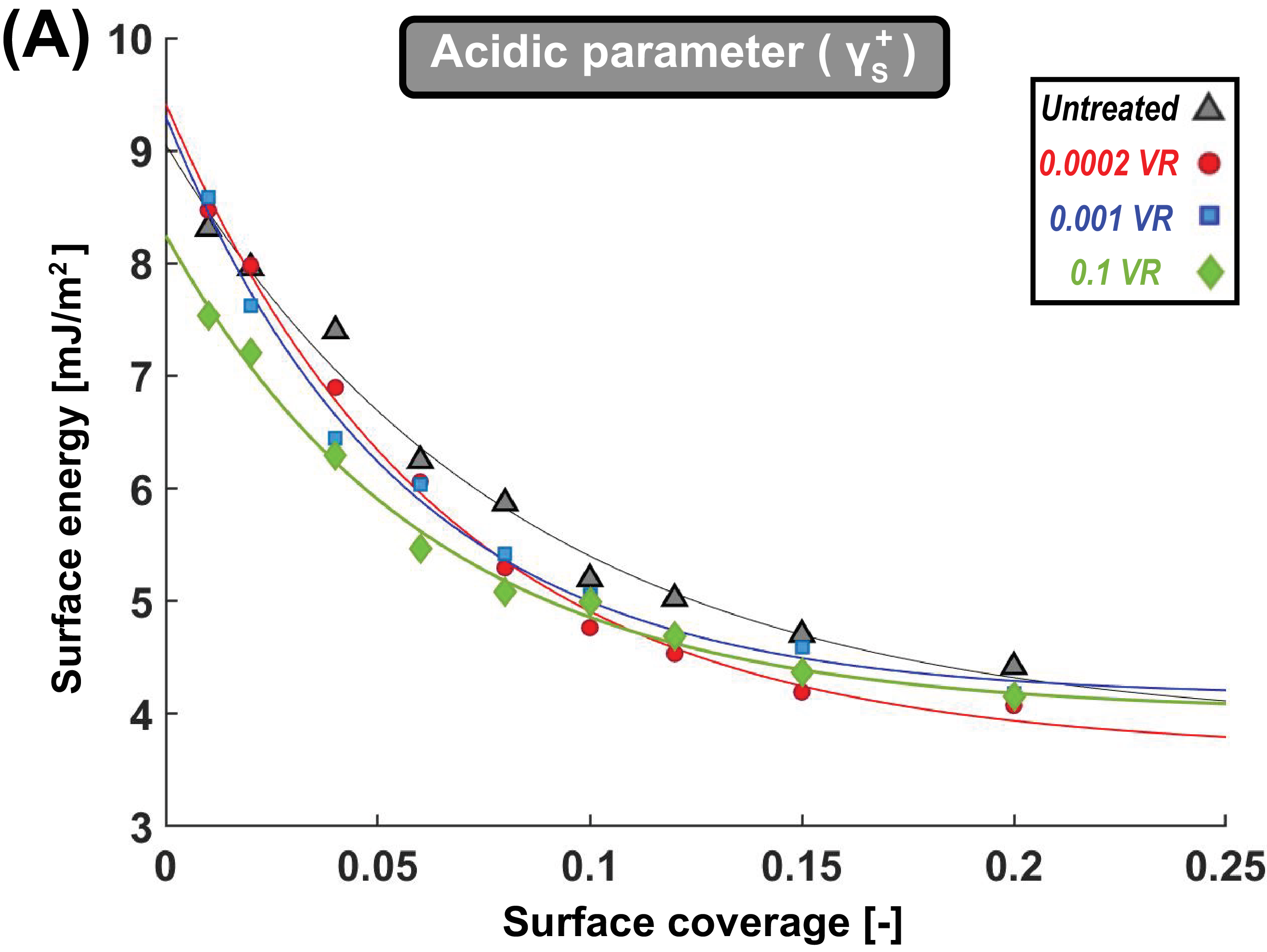}
        \end{subfigure}
        \hfill
        \begin{subfigure}[t]{0.48\textwidth}
            \centering
            \includegraphics[width=1\linewidth]{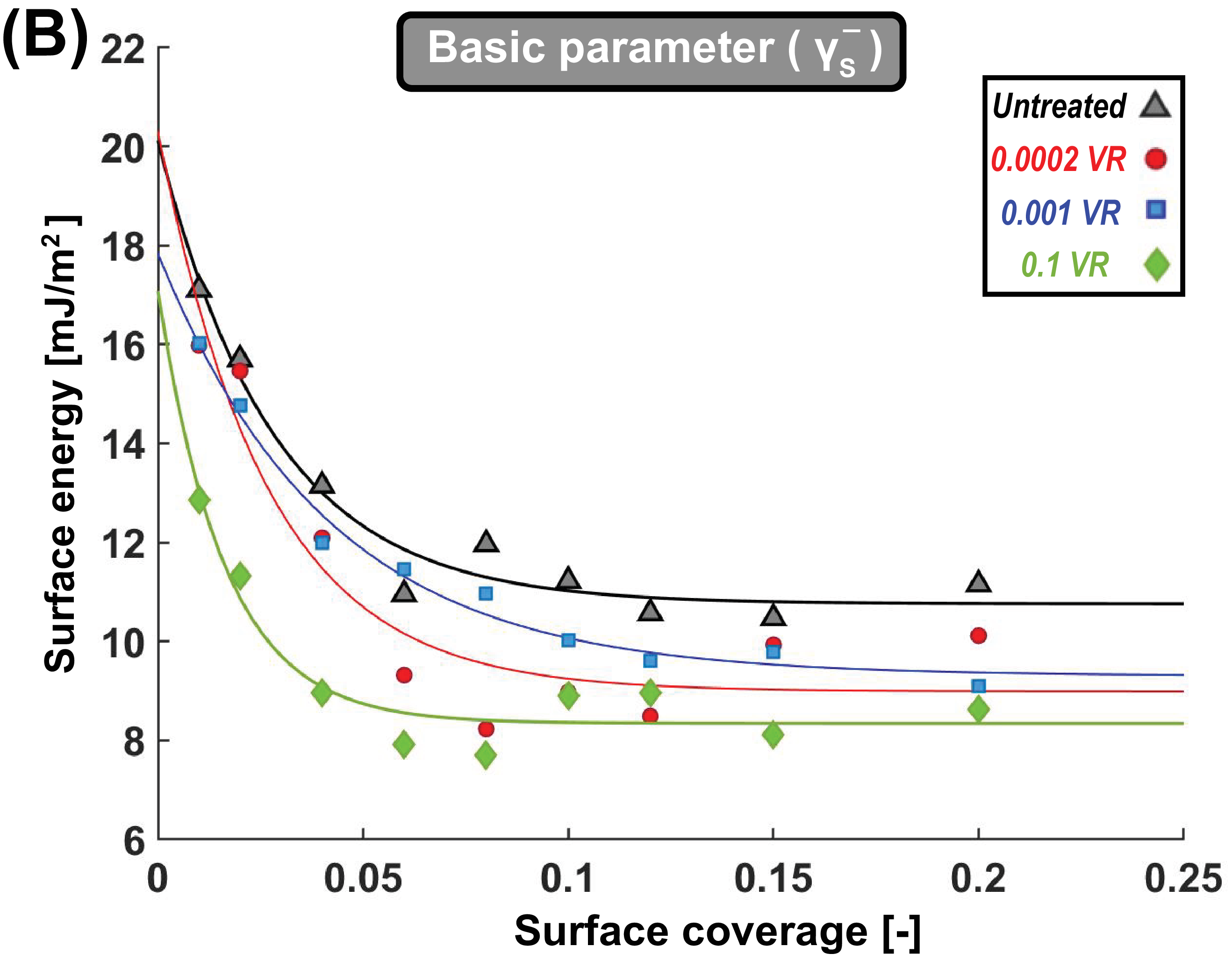}
        \end{subfigure}      
    \caption{Acidic and basic parameters of specific component of surface energy for untreated and treated glass beads.}
    \label{fig:Figure_6}
\end{figure}

Further insights into the nature of the chemical reaction between Surfasil and the glass surface can be derived from analyzing the acid and basic parameters of the specific component of surface energy. From an acid-base perspective, the silanol groups on the glass surface function as Lewis basic sites.  When Surfasil is applied as a silanizing agent, these basic sites are expected to be reduced, resulting in a lower basic parameter for the surface energy. This predicted decrease aligns with the results presented in \autoref{fig:Figure_6}, which shows a clear reduction in the basic parameter as the extent of silanization increases. In contrast, the acidic parameter remains nearly unchanged, indicating that Surfasil does not chemically interact with the acidic sites on the surface.

The silanization process not only reduces the specific surface energy components but also causes a reduction in the dispersive component, as shown in Figure \ref{fig:Figure_5}.a. The reduction in dispersive surface energy at high surface coverage, where the average energy of most surface sites is quantified (as explained in \autoref{Par:SE_distribution}), suggests that the added silanized layer likely has lower dispersive energy than the unmodified glass bead surface. Additionally, the reduction in dispersive surface energy at low surface coverage indicates that high-energy surface sites, such as cracks and narrow pores, participated in the chemical reaction during the silanization process. Given that the silanizing reagent acts as a coating, the involvement of these high-energy sites, particularly pores, may significantly influence the structural properties of the surface, such as its surface area. 

\begin{figure}[!ht]
    \centering
    \includegraphics[width=0.65\linewidth]{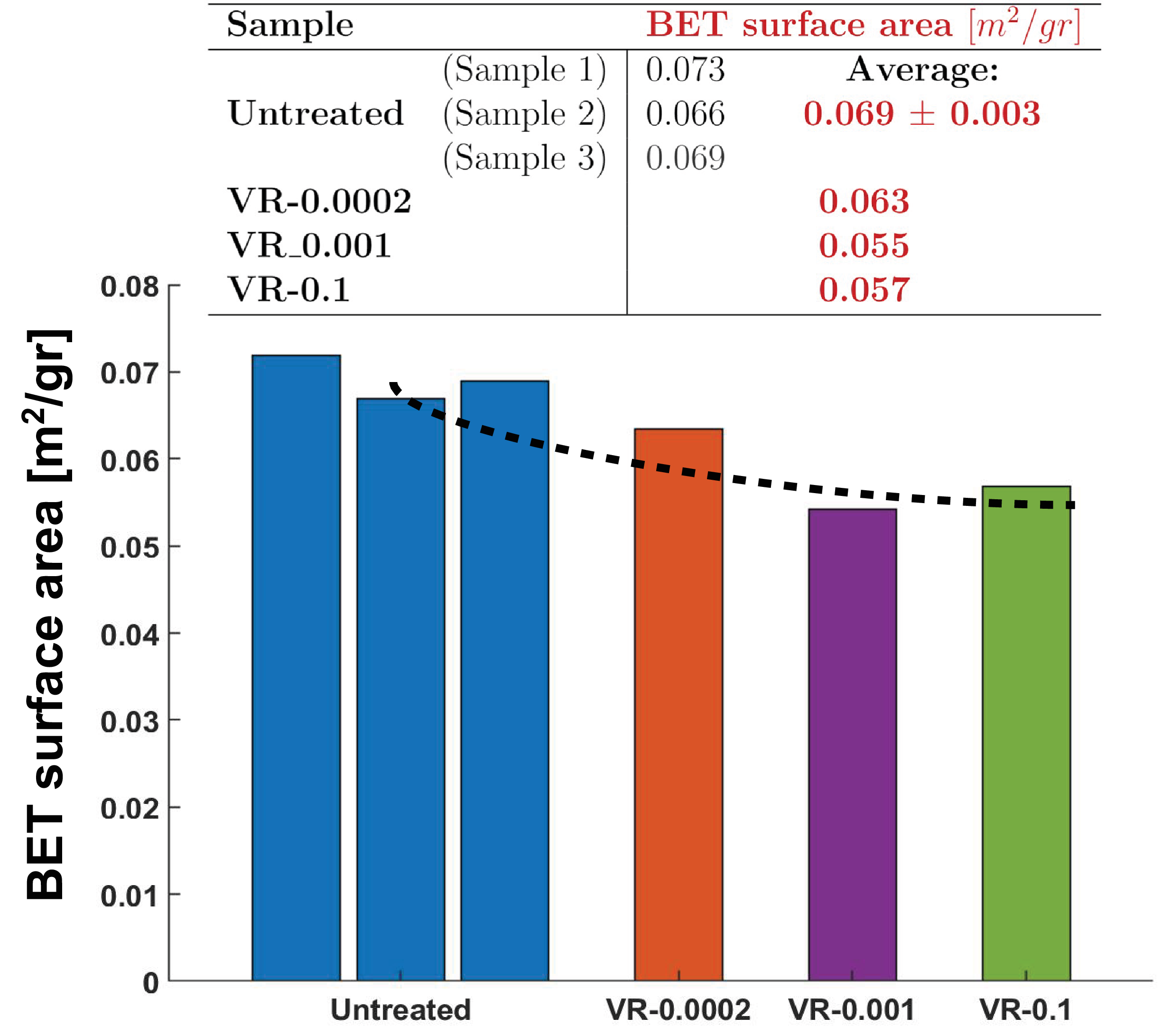}
    \caption{BET surface area of untreated (three replicates shown individually) and treated samples, calculated from n-Octane isotherms. A hand-drawn dashed curve highlights the decreasing trend with increasing silanization (not data-fitted).}
    \label{fig:Figure_7}
\end{figure}

To assess the structural modifications induced by silanization and further explore this possibility, the BET surface area and AFM images of the samples were analyzed. The BET surface area, shown in \autoref{fig:Figure_7}, decreases with increasing silanization, decreasing from an average of $0.0.069 \pm 0.003$ $m^2/g$ for the untreated glass to 0.057 $m^2/g$ for the VR-0.1 treated samples, indicating that the surface becomes progressively smoother as silanization progresses. This trend, which has also been observed in other studies \cite{Bauer2019FunctionalizationChromatography,Bauer2021SilanizationIGC}, highlights the reduction in BET surface area with silanization. Although the measured BET surface area is relatively small, it is important to note that IGC is reported to provide more accurate measurements for surface areas below 0.5 $m^2/g$ compared to the nitrogen adsorption technique \cite{Legras2015InverseArea}. The reduction in surface area at higher silanization concentrations appears to stabilize, suggesting that the surface area is reaching a plateau. This observation aligns with Vukovic’s findings, where further tuning of the treatment did not result in higher contact angles, indicating that the silanization process had reached the maximum packing density of the coating on the surface \cite{Vukovic2023SystematicModification}.

\begin{figure}[H]
    \centering
    \includegraphics[width=0.79\linewidth]{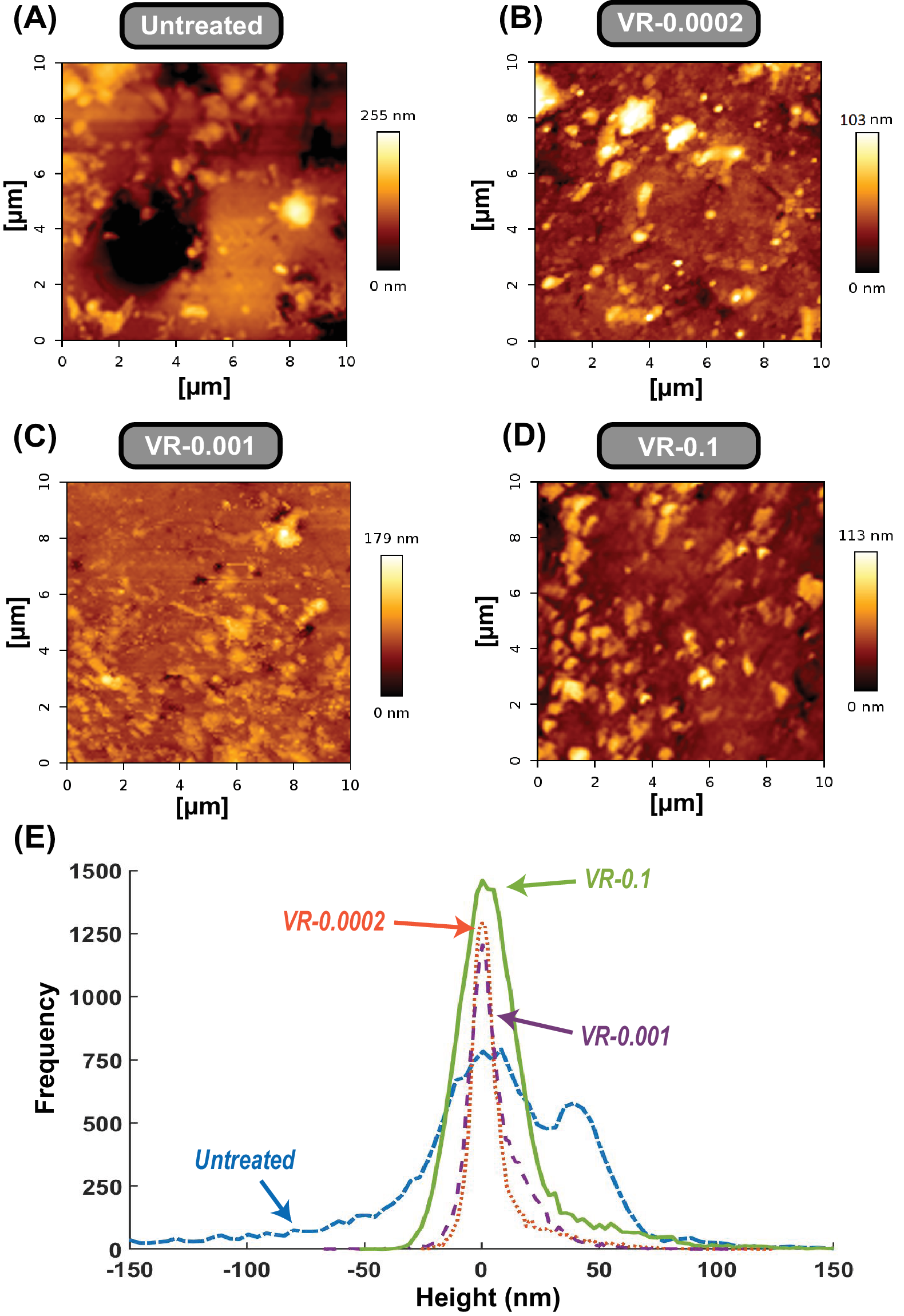}
    \caption{AFM Height Maps of Glass Bead Surfaces: (A) Untreated, (B) VR-0.0002, (C) VR-0.001, and (D) VR-0.1 treated samples. (E) Height histogram showing the distribution of surface feature heights. The histogram frequencies are plotted relative to the modal height, representing the most frequently observed height in each image.}
    \label{fig:Figure_8}
\end{figure}

Moreover, AFM  height maps, shown in Figure \ref{fig:Figure_8}.A-D, demonstrate that as silanization progresses, the large pores observed in the untreated glass diminish, and the overall height of surface features decreases. However, the number of surface features appears to increase, suggesting a patchy coverage of the silanized coating, consistent with previous findings \cite{Gaillard2018IsothermGlass}. The height histogram data in Figure \ref{fig:Figure_8}.E further supports these observations, showing a wide distribution of heights for the untreated sample compared to the narrower distribution in treated samples. This narrower distribution indicates that the height and depth of surface features are more uniform in treated samples, with smaller surface features compared to the untreated sample. These observations \textendash reduced BET surface area, the decreasing height of surface features in the AFM images, and the reduction in dispersive energy at low surface coverage \textendash collectively indicate that the surface becomes progressively smoother, yet more feature-dense, as a result of the silanization treatment.

\subsection{Surface Modifications of Samples from Water Adsorption During Storage} 

To investigate the stability of the coating and the extent of the structural and chemical modifications caused by water adsorption during storage, in this study, an untreated reference sample (sample column number 3) and a VR-0.1 silanized treated sample were subjected to 20 days of exposure to 90\% relative humidity, as outlined in \autoref{Sub_sub:IGC_exp_protocol}.

\begin{figure}[!ht]
        \centering
        \begin{subfigure}[t]{0.48\textwidth}
            \centering
            \includegraphics[width=1\linewidth]{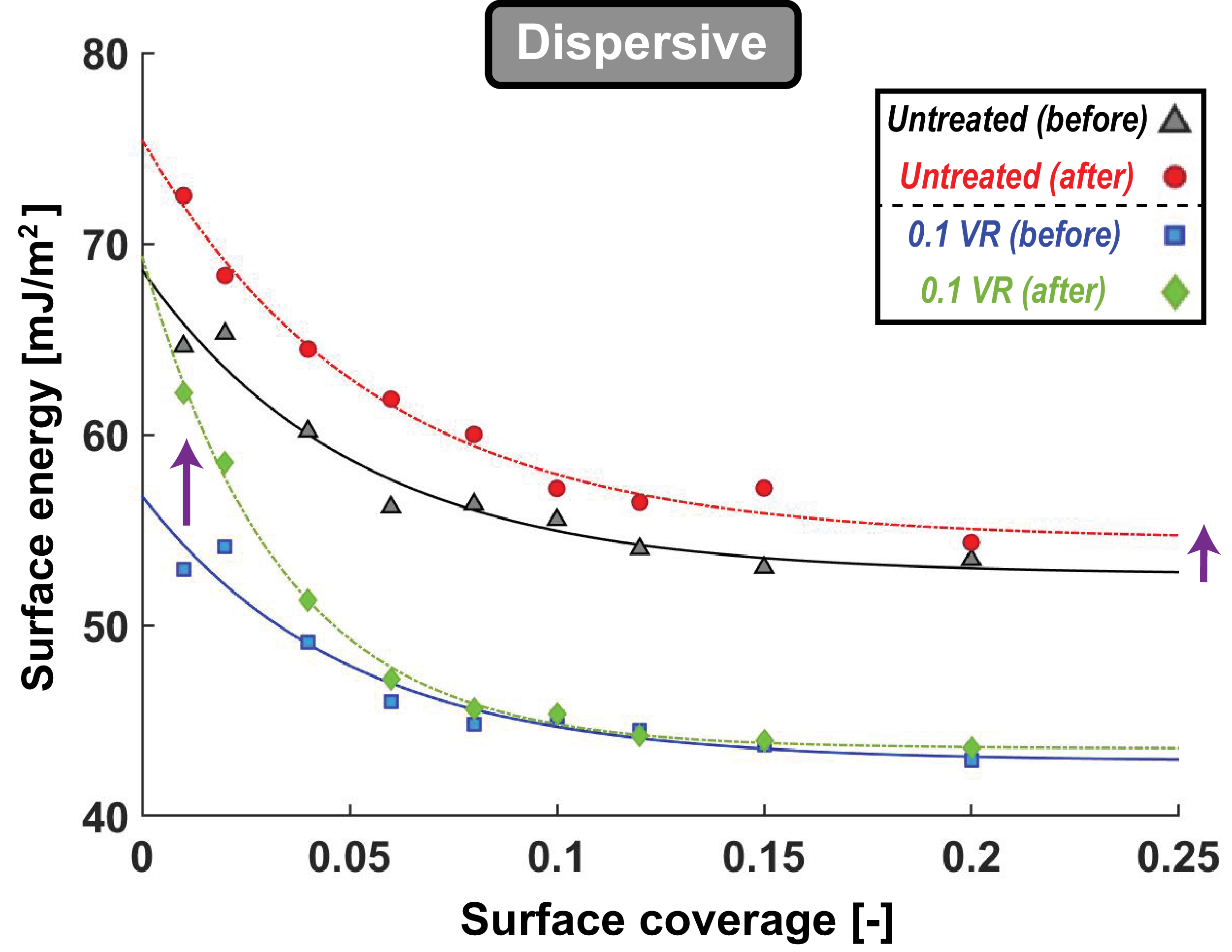}
        \end{subfigure}
        \hfill
        \begin{subfigure}[t]{0.48\textwidth}
            \centering
            \includegraphics[width=1\linewidth]{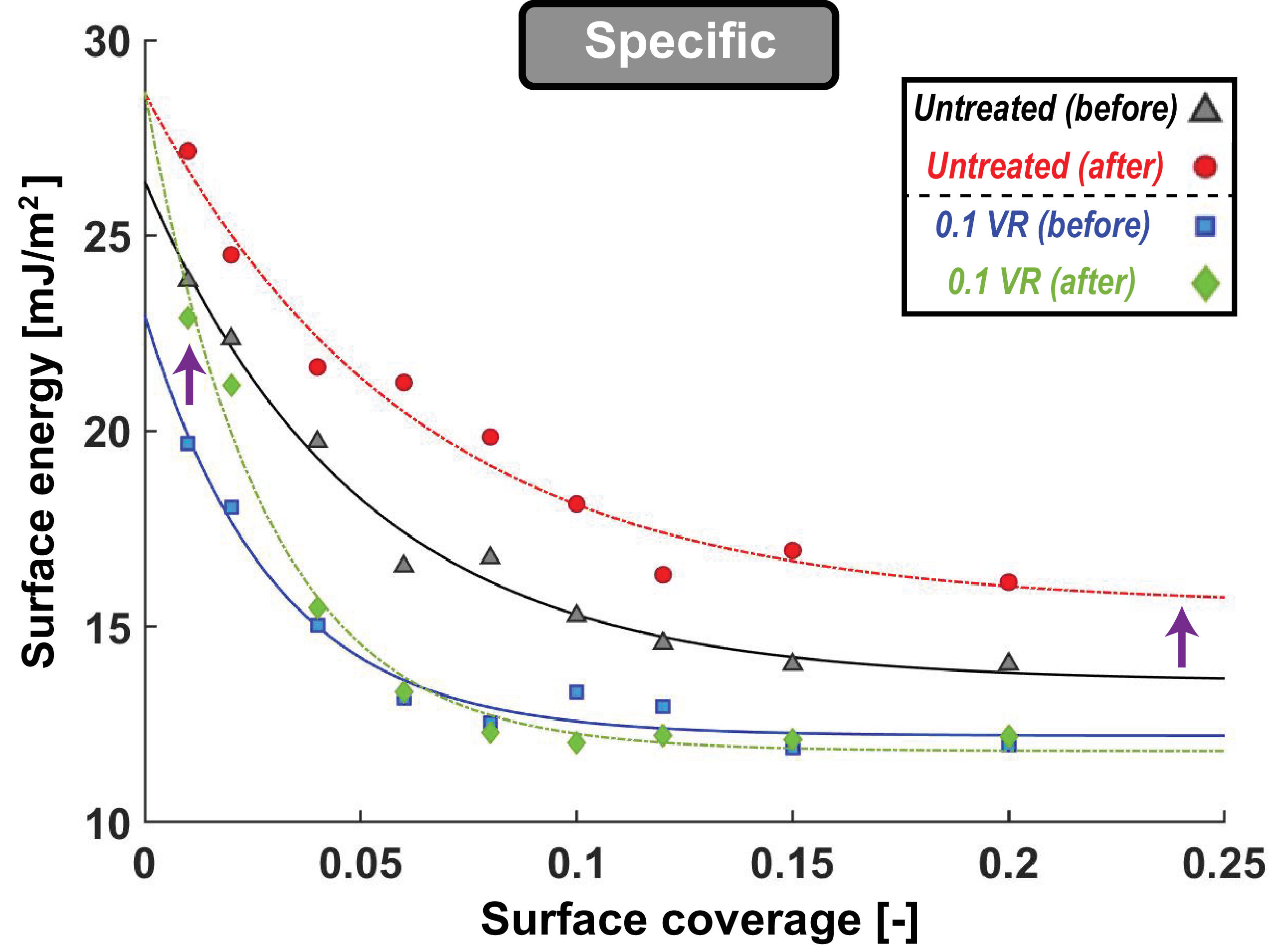}
        \end{subfigure}      
    \caption{Dispersive and specific surface energy components of untreated and VR-0.1 silanized glass bead samples, measured before and after exposure to 90\% relative humidity to simulate long-term storage.}
    \label{fig:Figure_9}
\end{figure}

The dispersive and specific components of surface energy, measured before and after exposure to relative humidity, are presented in Figure \ref{fig:Figure_9}. The untreated glass bead sample exhibits an increase in both components across entire surface coverage, suggesting that water adsorption affected the majority of the surface. In contrast, the VR-0.1 treated sample shows an increase in these components only at low surface coverage, indicating that modifications caused by adsorbed water were limited to high-energy sites. In both samples, the increase in the specific component of surface energy is attributed to the reaction of adsorbed water with the surface, leading to the formation of additional polar groups, as depicted in Figure \ref{fig:Figure_1} \cite{Yeon2023HydroxylationSimulations,Gin2021AqueousPerspectives}.

The reaction of adsorbed water with the glass surface is also expected to cause surface roughening \cite{Gin2021AqueousPerspectives}. Given that the reaction primarily generates polar groups, the observed increase in the dispersive component of the surface energy of samples is likely attributed to this roughening. To further explore these structural modifications, BET surface area measurements and AFM images of the samples were analyzed. The BET surface area of the untreated glass beads increased from $0.069$ $m^2/gr$ to 0.078 $m^2/gr$ following exposure to relative humidity, while the treated sample showed only a marginal increase, from 0.057 $m^2/gr$ to 0.062 $m^2/gr$, indicating that the silanized coating largely resisted the effects of water adsorption. The AFM images, taken over a $10 \mu m \times 10 \mu m$ area and provided in the \ref{Sec:Supporting_Information}, revealed small-scale changes that further support this finding, suggesting that roughness-induced alterations occur at a nanoscale.

\begin{figure}[!ht]
    \centering
    \includegraphics[width=0.8\linewidth]{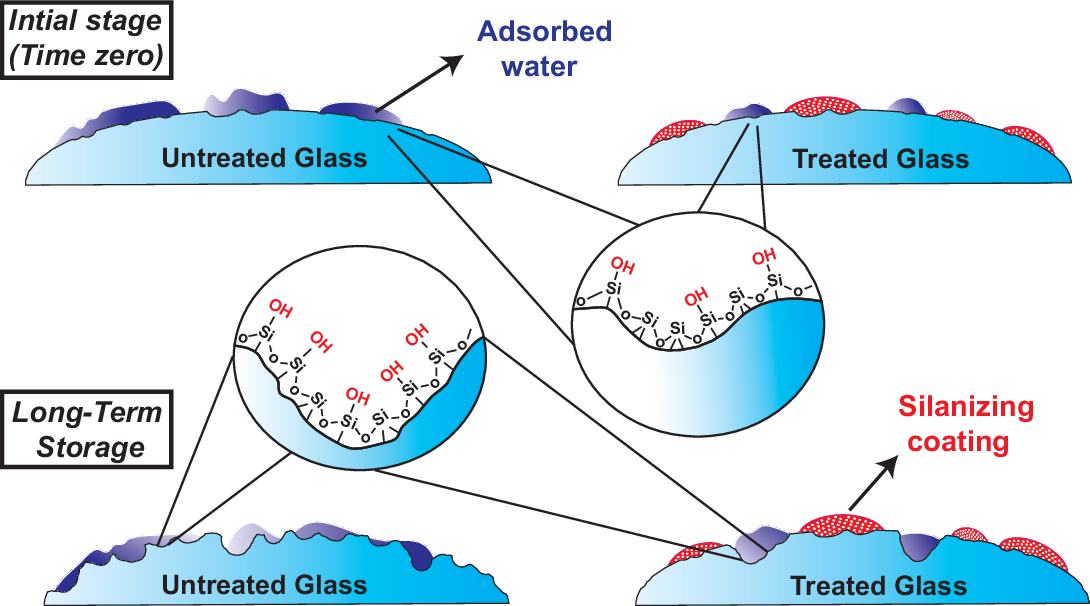}
    \caption{Schematic of surface modifications on untreated and silanized soda lime glass beads. "Initial" shows time zero, and "Long-Term" illustrates the effects of water adsorption after extended storage.}
    \label{fig:Figure_10}
\end{figure}

The results observed up to this point suggest that the silanized coating is unlikely to engage in chemical reactions with adsorbed water during storage. The patchy coverage of the coating \cite{Gaillard2018IsothermGlass}, combined with its hydrophobic nature, supports this observation, as water molecules from the gas phase preferentially adsorb onto the exposed regions of the glass surface that are not covered by the coating. Based on the insights gained from the results, \autoref{fig:Figure_10} illustrates a conceptual model developed to explain the distinct surface modifications observed in untreated and silanized glass during storage under humid conditions, driven by water adsorption. In the treated sample, the adsorbed water is shown to form primarily on the uncovered regions of the silanized coating, consistent with the hydrophobic nature of the coating. The structural and chemical changes caused by adsorbed water, as depicted in the model after long-term storage, occur in these specific regions where water was initially adsorbed. The minimal extent of these changes aligns with the slight increase in BET surface area and the increase in dispersive and specific surface energy components observed at low surface coverage. In contrast, in the untreated glass bead sample, water adsorption is not confined to specific regions. The relatively greater extent of chemical and structural changes shown after long-term storage corresponds to a broader increase in BET surface area and surface energy components over the full range of surface coverage. In both samples, the chemical changes are represented by the formation of additional hydroxyl groups on the surface.

It is essential to note that the conceptual model was developed based on experimental observations from untreated and silanized soda lime glass beads, without accounting for the influence of atoms or ions from salts incorporated into the glass matrix during manufacturing \cite{Shelby2020IntroductionTechnology}. The extent and rate of nanoscale chemical and structural modifications are highly dependent on the glass's chemical composition and dissolution of these incorporated atoms or ions of salts can intensify the surface modifications \cite{Yeon2023HydroxylationSimulations,Gin2021AqueousPerspectives}. Nevertheless, nanoscale structural and chemical modifications are likely to occur on any type of glass surface. These surface alterations can significantly impact the distribution of thin fluid films, which play a crucial role in regulating multiphase flow behavior, particularly near the saturation point of the wetting fluid \cite{Armstrong2021MultiscaleMedia,Wensink2024In-situRocks}. As a result, the interaction between adsorbed water from air and the glass surface during storage can lead to variations in surface properties, potentially affecting multiphase flow patterns and causing inconsistencies in repeated experiments.

\section{Conclusions}

This study investigated the nanoscale structural and chemical modifications of glass bead surfaces with varying wettability, achieved through a silanization procedure developed by Vukovic et al. \cite{Vukovic2023SystematicModification}. The focus was on providing detailed nanoscale analysis of the structural and chemical changes induced by the silanization process, expanding beyond the contact angle measurements employed in the original work. Additionally, the study examined the stability of the treated glass bead samples during storage, evaluating how their structural and chemical properties change over time due to reactions with adsorbed water.

Silanization led to considerable nanoscale modifications to the structural and chemical properties of the glass bead surfaces, resulting in a reduction in hydrophilicity. This was evident from increased contact angles and a corresponding decrease in both dispersive and specific surface energy components. Analysis of the specific component of surface energy revealed that the silanizing agent selectively targeted basic polar groups, such as silanol groups, as indicated by the decrease in the basic parameter of surface energy, while the acidic parameter remained largely unchanged.

BET surface area measurements and AFM images indicated that the surface became progressively smoother with increased silanization, likely due to the silanizing agent covering high-energy features such as pores and cracks. This conclusion is supported by the observed decrease in the dispersive component of surface energy at low surface coverage, suggesting that these high-energy sites were effectively involved in the silanization process. Furthermore, the decrease in dispersive energy at high surface coverage, where the average energy of most surface sites is quantified, suggests that the silanized layer exhibited lower energy than the untreated glass surface.  AFM images also revealed that, while the large pores present in the untreated glass diminished with silanization and the overall height of surface features decreased, the number of these surface features increased, indicating a patchy coverage \cite{Gaillard2018IsothermGlass}. These findings confirm that silanization not only selectively targeted specific chemical groups but also induced structural changes that resulted in a smoother surface yet more feature-dense surface. 

The second part of the study assessed the stability of the treated sample and the extent of its structural and chemical modifications induced by water adsorption during storage, using untreated glass beads as a reference. The findings demonstrated that the silanized coating on the VR-0.1 treated sample provided strong resistance to water vapor, with minimal changes observed in both BET surface area and surface energy components after 20 days exposure to 90\% relative humidity, a condition used to accelerate reactions typically occurring during long-term storage. The treated sample exhibited slight increases in surface energy at low coverage, suggesting that only localized modifications in high-energy areas, such as small pores and cracks, had occurred.

In contrast, the untreated glass bead sample underwent more extensive surface modifications after exposure, as indicated by the increased BET surface area and notable changes in both dispersive and specific surface energy components. This suggests that water adsorption led to surface roughening and the formation of additional polar groups, as illustrated in Figure \ref{fig:Figure_1}. Based on the insights gained from the results, a conceptual model (Figure \ref{fig:Figure_10}) was developed to illustrate the distinct changes in untreated and silanized glass surfaces during storage, highlighting the effects of adsorbed water reactions over time. Although the extent of these structural and chemical modifications depends on storage conditions and duration, they can impact the distribution and behavior of thin films on the surface. This, in turn, may lead to inconsistencies in multiphase flow behavior during repeated experiments, particularly near the saturation endpoint of the wetting fluid \cite{Armstrong2021MultiscaleMedia,Wensink2024In-situRocks}. 

\section*{CRediT authorship contribution statement}
\textbf{M.H.Khoeini:} Writing - Original Draft, Conceptualization, Methodology, Investigation, Formal analysis, Validation, Software, Visualization, Resources.
\textbf{G. Wensink:} Writing - Review \& Editing, Investigation, Formal analysis.
\textbf{T.Vukovic:} Investigation, Formal analysis, Resources.
\textbf{I.Krafft:} Visualization, Formal analysis.
\textbf{A.Van der Net:} Writing - Review \& Editing, Supervision, Funding acquisition. 
\textbf{M.R\"ucker:} Writing - Review \& Editing, Supervision, Funding acquisition, Project administration.
\textbf{A. Luna-Triguero:} Writing - Review \& Editing, Conceptualization, Supervision, Project administration.

\section*{Declaration of Competing Interest}
The authors declare that they have no known competing financial interests or personal relationships that could have appeared to influence the work reported in this paper.

\section*{Acknowledgment}
This research has received funding from the NWO DeepNL program (DEEP.NL.2019.006) and NWO (project 19121). The Research Council of Norway is acknowledged for its support through project 262644 (Pore-lab Center of Excellence).

\appendix
\section{AFM Images of Untreated and Treated Samples Pre- and Post-Exposure to 90\% Relative Humidity.}
\label{Sec:Supporting_Information}
\setcounter{figure}{0} 

\begin{figure}[!ht]
    \centering
    \includegraphics[width=0.9\linewidth]{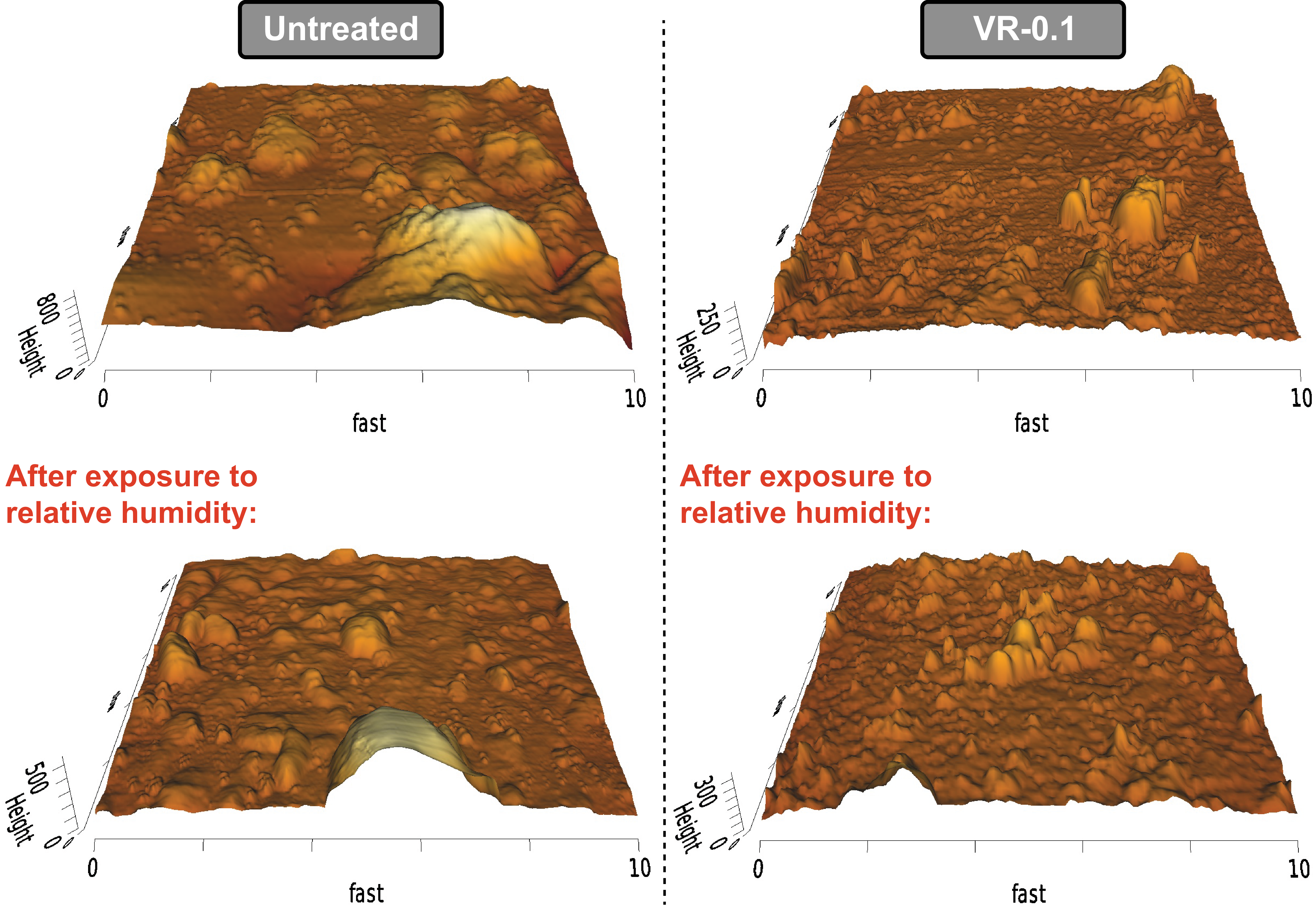}
    \caption{AFM images of untreated (left) and VR-0.1 silanized treated (right) glass bead samples before (top) and after (bottom) exposure to 90\% relative humidity for 20 days.}
    \label{fig:Figure_A1}
\end{figure}

AFM images of untreated and VR-0.1 silanized treated samples, taken before and after exposure to 90\% relative humidity for 20 days, are presented in \autoref{fig:Figure_A1}. In the untreated sample (left), a visible increase in small-scale surface features is evident after humidity exposure, with more pronounced peaks and valleys in the bottom image compared to the top. This indicates that the untreated glass surface undergoes increased surface roughening due to moisture adsorption, likely resulting from the formation of additional polar groups and structural changes. In contrast, the VR-0.1 silanized samples (right) exhibit less pronounced changes in surface topography after humidity exposure, indicating that the silanized coating provides some resistance to moisture-induced roughening. However, subtle alterations in surface features can still be observed in the bottom image. These small-scale changes suggest that finer details of surface modifications are present, though not fully captured at the current scan resolution.

\bibliography{references}
\bibliographystyle{elsarticle-num}

\end{document}